\documentclass[a4paper]{jpconf}
\usepackage{graphicx,url,amsmath,amssymb}
\usepackage{hyperref}
\newcommand{\pd}[3]{\frac{\partial^{#3} #1}{\partial {#2}^{#3}}} 
\newcommand{\td}[3]{\frac{\mathrm{d}^{#3} #1}{\mathrm{d} {#2}^{#3}}} 
\urldef{\footurlone}\url{https://apps.sarao.ac.za/calculators/continuum}
\newcommand{\footurl}[1]{\footnote{\url{#1}}}

\begin{document}
\title{Just a MeerKAT, or a dark matter machine?}
\author{G Beck and S Makhathini}
\address{School of Physics and Centre for Astrophysics, University of the Witwatersrand,
Johannesburg, Wits 2050, South Africa.}
\ead{geoffrey.beck@wits.ac.za,sphesihle.makhathini@wits.ac.za}

\begin{abstract}
    The MeerKAT telescope is a precursor experiment to the full Square Kilometre Array. The latter's potential to explore the nature of dark matter, via indirect detection, has received attention previously in the literature. In this work, we demonstrate the potential of MeerKAT to make initial forays into the dark matter parameter space through a {\tt stimela}-based simulation framework. In particular, we show that 20~hr MeerKAT observations at U- and L-band of the dwarf galaxy Reticulum II can improve the constraints on WIMP dark matter by a factor of 3 compared to current gamma-ray observations. Furthermore, these MeerKAT constraints are an order magnitude better than Fermi-LAT estimates from similar galaxies. 
    
\end{abstract}

\section{Introduction}

Radio frequency searches for Dark Matter (DM) have long been a minor consideration compared to gamma-ray searches. This is due to the presence of larger radio backgrounds as well as the fact that synchrotron emissions require the characterisation of magnetic field environment in cosmic structures. However, with recent advances in radio interferometry techniques and software tools, these challenges can start to be addressed. More specifically, tools like {\tt Cubical}, {\tt KillMS}, {\tt WSClean}, {\tt DDFacet}, and {\tt PyBDSF} \cite{Kenyon_2018,killms1,killms2,offringa-wsclean-2014,ddfacet,pybdsf} are sophisticated enough to do high fidelity background subtraction and full-stokes calibration required to characterize magnetic fields. In this paper, we consider the potential of SKA precursor, MeerKAT, to make initial strides in a radio-frequency hunt for DM, before the full SKA instrument is available. 

The potential of radio probes, for DM indirect detection, has been well studied in the literature~\cite{baltz1999,baltz2004,Colafrancesco2006,Colafrancesco2007,siffert2011,gsp2015,storm2017}. This work has led to several empirical follow-ups in dwarf galaxies producing non-detection limits~\cite{natarajan2013,spekkens2013,atca_III,Regis_2017,Vollmann_2020,Basu_2021,Regis_2021} that have been competitive with, and sometimes exceed, those from Fermi-LAT~\cite{fermi} in similar targets. The observational works have made use of several telescopes: LOFAR, GMRT, ATCA, and ASKAP. However, the South African MeerKAT telescope is yet to step into this fray. MeerKAT boasts the best sensitivity of any of these telescopes\footnote{\footurlone} between 540 and 1650~MHz and should lead to the most stringent limits on Weakly Interacting Massive Particles (WIMPs) at these frequencies. In this work, we will examine sensitivity estimates for the MeerKAT array using the {\tt stimela} pipe-lining package~\cite{makhathini2018}. In particular, to explore the effect of a visibility taper\footnote{A visibility taper down-weights visibilities from long baselines which leads to lower image resolution. The `target' resolution is a tunable parameter} on the potential non-detection constraints in the dwarf galaxy Reticulum II. This is due to the fact that the use of the visibility taper is designed to focus on large scale diffuse emissions and we explore a tapering scale of $15^{\prime\prime}$ to increase our sensitivity to the expected arcminute-scale diffuse emission from DM in dwarf galaxies. 

The galaxy Reticulum II is a highly suitable target for southern hemisphere observation due to both its DM density and declination~\cite{Pace_2018,Koposov_2015}. The former offers stronger indirect DM signals while the latter ensures lower radio backgrounds due to the distance from the galactic plane. For this work, we assume a model-independent WIMP dark matter signal. Our findings indicate  strong non-detection upper limits on the WIMP annihilation cross-section in $b\bar{b}$, $\tau^+\tau^-$, and $\mu^+\mu^-$. These improved constraints will allow us to probe below the thermal relic value~\cite{steigman2012} for WIMP masses below 300~GeV with a minimal magnetic field strength of $0.5$~$\mu$G, close to the expected minimum in dwarf galaxies from star formation arguments~\cite{atca_II}. With a larger field strength of 1~$\mu$G, MeerKAT has the potential to rule out WIMPs annihilating via $b$-quarks when the mass is below 1 TeV with just 20 hours of observation of Reticulum II.

The rest of the paper is structured as follows: in Section~\ref{sec:radio} we lay out the formalism for DM synchrotron emissions, while in Section~\ref{sec:meerKAT} we detail our method for determining the sensitivity of the MeerKAT telescope to diffuse radio emissions. Finally, results are presented in Section~\ref{sec:results} and discussed in Section~\ref{sec:disc}.

\section{Radio emission from dark matter}
\label{sec:radio}
The synchrotron emission from electrons/positrons that result from the annihilation or decay of DM can be determined by solving the diffusion-loss equation
\begin{equation}
\pd{}{t}{}\td{n_{\mathrm{e}}}{E}{} =  \vec{\nabla} \left( D(E,\vec{r})\vec{\nabla}\td{n_{\mathrm{e}}}{E}{}\right) + \pd{}{E}{}\left( b(E,\vec{r}) \td{n_{\mathrm{e}}}{E}{}\right) + Q_{\mathrm{e}}(E,\vec{r}) \; ,\label{eq:diff-loss}
\end{equation}
where $D(E,\vec{r})$ is the diffusion function, $b(E,\vec{r})$ the energy-loss function and $\td{n_{\mathrm{e}}}{E}{}$ is the electron distribution. The source function $Q_{\mathrm{e}}$ is given by
\begin{equation}
Q_{\mathrm{e,f}} (r,E) = \frac{1}{2}\langle \sigma V\rangle  \td{N^{\mathrm{f}}_{\mathrm{e}}}{E}{} \left(\frac{\rho_{\chi}(r)}{m_{\chi}}\right)^2\; ,
\end{equation}
where $\rho_{\chi}$ is the DM density distribution, $m_\chi$ is the DM mass, $f$ is the annihilation/decay channel, $\td{N^{\mathrm{f}}_{\mathrm{e}}}{E}{}$ is the injected spectrum from channel $f$, $\langle \sigma V\rangle$ is the cross-section for annihilation, and $\Gamma$ is the decay rate. 

A Green's function can used to find the stationary solutions to Eq~(\ref{eq:diff-loss}), when $D$ and $b$ have no $\vec{r}$ dependence~\cite{baltz1999,baltz2004,Colafrancesco2006,Colafrancesco2007} and we follow the prescription from \cite{Colafrancesco2006} for our solution. The stationary solutions can then be used to determine the emissitivity as 

\begin{equation}
j_{\mathrm{sync}} (\nu,r,z) = \int_{m_{\mathrm{e}}}^{m_\chi} dE \, \left(\td{n_{\mathrm{e}^-}}{E}{} + \td{n_{\mathrm{e}^+}}{E}{}\right) P_{\mathrm{sync}} (\nu,E,r,z) \; ,
\label{eq:emm}
\end{equation}

where $P_{\mathrm{sync}}$ is the synchrotron power emitted at frequency $\nu$ by an electron with energy $E$ at redshift $z$ and position $\vec{r}$, and is being given by~\cite{rybicki1986,longair1994}
\begin{equation}
P_{\mathrm{sync}} (\nu,E,r,z) = \int_0^\pi d\theta \, \frac{\sin{\theta}}{2}2\pi \sqrt{3} r_{\mathrm{e}} m_{\mathrm{e}} c \nu_{\mathrm{g}} F_{\mathrm{sync}}\left(\frac{\kappa}{\sin{\theta}}\right) \; ,
\label{eq:power}
\end{equation}
here $r_{\mathrm{e}}$ is the classical electron radius and $\nu_g = \frac{e B}{2\pi m_{\mathrm{e}} c}$ is the non-relativistic gyro frequency. The values of $\kappa$ and $F_{\mathrm{sync}}$ are given by
\begin{equation}
\kappa = \frac{2\nu (1+z)}{3\nu_{\mathrm{g}} \gamma^2}\left[1 +\left(\frac{\gamma \nu_\mathrm{p}}{\nu (1+z)}\right)^2\right]^{\frac{3}{2}} \; ,
\end{equation}
and
\begin{equation}
F_{\mathrm{sync}}(x) = x \int_x^{\infty} dy \, K_{5/3}(y) \approx 1.25 x^{\frac{1}{3}} \mbox{e}^{-x} \left(648 + x^2\right)^{\frac{1}{12}} \; .
\end{equation}

Finally, we can define the surface brightness at cylindrical radius $r$ via
\begin{equation}
I_{\mathrm{sync}} (\nu,r,z) = \int_r^\infty d r^{\prime} \, \frac{r^\prime j_{\mathrm{sync}}(\nu,r^{\prime},z)}{4 \pi \sqrt{\left(r^{\prime}\right)^2-r^2}} \; ,
\label{eq:fluxsb}
\end{equation}
where the integration runs over the spherical radius $r^\prime$. 

\section{Emission in Reticulum II}
Besides its favourable location, with respect to the galactic plane, and its high DM density environment, Reticulum II has already been studied as a candidate for indirect DM detection using observations from ATCA~\cite{Regis_2017}. In fact, this previous study gives us a model to characterise the radio sky in our region of interest (see Section \ref{sec:meerKAT}).

The radial DM density profile will be taken as an Einasto halo~\cite{einasto1968} following \cite{Regis_2017}
\begin{equation}
    \rho (r) = \rho_\mathrm{s} \exp\left[-\frac{2}{\alpha} \left(\left[\frac{r}{r_s}\right]^{\alpha} - 1\right)\right] \; ,
\end{equation}
where $\alpha$ is the profile slope index, as well as $\rho_\mathrm{s}$ and $r_s$ being the characteristic density and radius. Note that the various parameter values given in Table~\ref{tab:ret2}. The choice of $\alpha$ is motivated by studies arguing for cored profiles in dwarf galaxies~\cite{walker2009,adams2014}.
We will assume the magnetic field has an exponential profile given by~\cite{Regis_2017}
\begin{equation}
    B(r) = B_0 \; \exp{\left(-\frac{r}{r_*}\right)} \; .
\end{equation}
It is argued in \cite{atca_II} that dwarf galaxies can be expected to have a peak magnetic field strength $B_0 \gtrsim 0.4$ $\mu$G and that the Milky-Way magnetic field strength is $\approx 1.4$ $\mu$G at the location of Reticulum II. Given the lack of observational limits on $B_0$ in dwarf galaxies~\cite{atca_II}, we will explore a range of $0.5$ to $2$ $\mu$G as possible field strength values.

The thermal electron density is similarly given by
\begin{equation}
    n (r) = 10^{-6} \, \mathrm{cm}^{-3} \; \exp{\left(-\frac{r}{r_*}\right)} \; ,
\end{equation}
where $r_*$ is the stellar half-light radius (see Table~\ref{tab:ret2} for the value). When computing $\bar{n}$ and $\bar{B}$, for use in Green's functions, we will weight the averages according to $\rho^2$, in order to better represent the environment where the annihilations take place.

\begin{table}[ht!]
    \centering
    \begin{tabular}{|c|c|c|}
         \hline
         Property & Value & Ref \\
         \hline
         $r_*$ & 35 pc & \cite{Koposov_2015}\\
         $\rho_\mathrm{s}$ & $7\times 10^{16}$ M$_\odot$ Mpc$^{-3}$ & \cite{Regis_2017} \\
         $r_\mathrm{s}$ & 200 pc & \cite{Regis_2017} \\
         $\alpha$ & 0.4 & \cite{Regis_2017}\\
         $D_\mathrm{L}$ & 30 kpc & \cite{Koposov_2015} \\
         \hline
    \end{tabular}
    \caption{Reticulum II dwarf galaxy properties.}
    \label{tab:ret2}
\end{table}

\section{Simulating MeerKAT's sensitivity}
\label{sec:meerKAT}
We use {\tt stimela} to produce realistic MeerKAT simulations. This allows us to assemble an end-to-end sky and telescope simulation that follows this process:

\begin{enumerate}
    \item Use the {\tt simms} package \footurl{https://github.com/ratt-ru/simms} (based on {\tt CASA} \cite{casa}) to simulate a MeerKAT visibility data set. 
    \item Populate the simulated visibility with those from a sky model of Reticulum II, found in \cite{Regis_2017}, using {\tt MeqTrees}~\cite{meqtrees}. Direction-independent antenna gains as well as thermal noise are also added at this step.
    \item Use the {\tt RFIMasker}\footurl{https://github.com/bennahugo/RFIMasker} tool to flag data at frequencies known to be contaminated by radio frequency interference (RFI) as unsuable. This flags around 10\% of the data between 540 and 1080~MHz (U-band) and 36\% between 890 and 1650~MHz (L-band)\footurl{https://skaafrica.atlassian.net/wiki/spaces/ESDKB/pages/305332225/Radio+Frequency+Interference+RFI}. 
    \item Calibrate the data to correct for the direction-independent gains using {\tt MeqTrees}. 
    \item The data are then imaged and deconvolved using {\tt WSClean}.
\end{enumerate}

We use a channel width of 10~MHz and an integration time of 8 seconds for both the U-band L-band. The total synthesis time is taken to be 20 hours. For the L-band we find an rms noise of $4$ $\mu$Jy/beam with a beam FHWM of $9.1^{\prime\prime} \times 6.4^{\prime\prime}$, while the U-band simulations we have $8.3$~$\mu$Jy/beam and with a beam of FWHM $13.9^{\prime\prime} \times 9.8^{\prime\prime}$. The quoted rms and beam values are from imaging with Briggs $uv$-weighting with a robust parameter of zero. Now, these resolutions (beam FWHM) are not suitable for detecting diffuse emission at our target angular scales. Therefore, as was done in \cite{Regis_2017}, we use a visibility taper of 15 arcseconds since it produces a resolution that better matches our target scales. With this taper, we estimate the MeerKAT L-band sensitivity at $1.7$~$\mu$Jy/beam with a beam FWHM of $19.8^{\prime\prime} \times 16.5^{\prime\prime}$ and a U-band sensitivity of $4.4$~$\mu$Jy/beam with a beam FWHM of $22.0^{\prime\prime} \times 17.2^{\prime\prime}$. This represents an order of magnitude improvement in point-source sensitivity.

\section{Results}
\label{sec:results}

\begin{figure}[ht]
    \centering
    \resizebox{0.48\hsize}{!}{\includegraphics{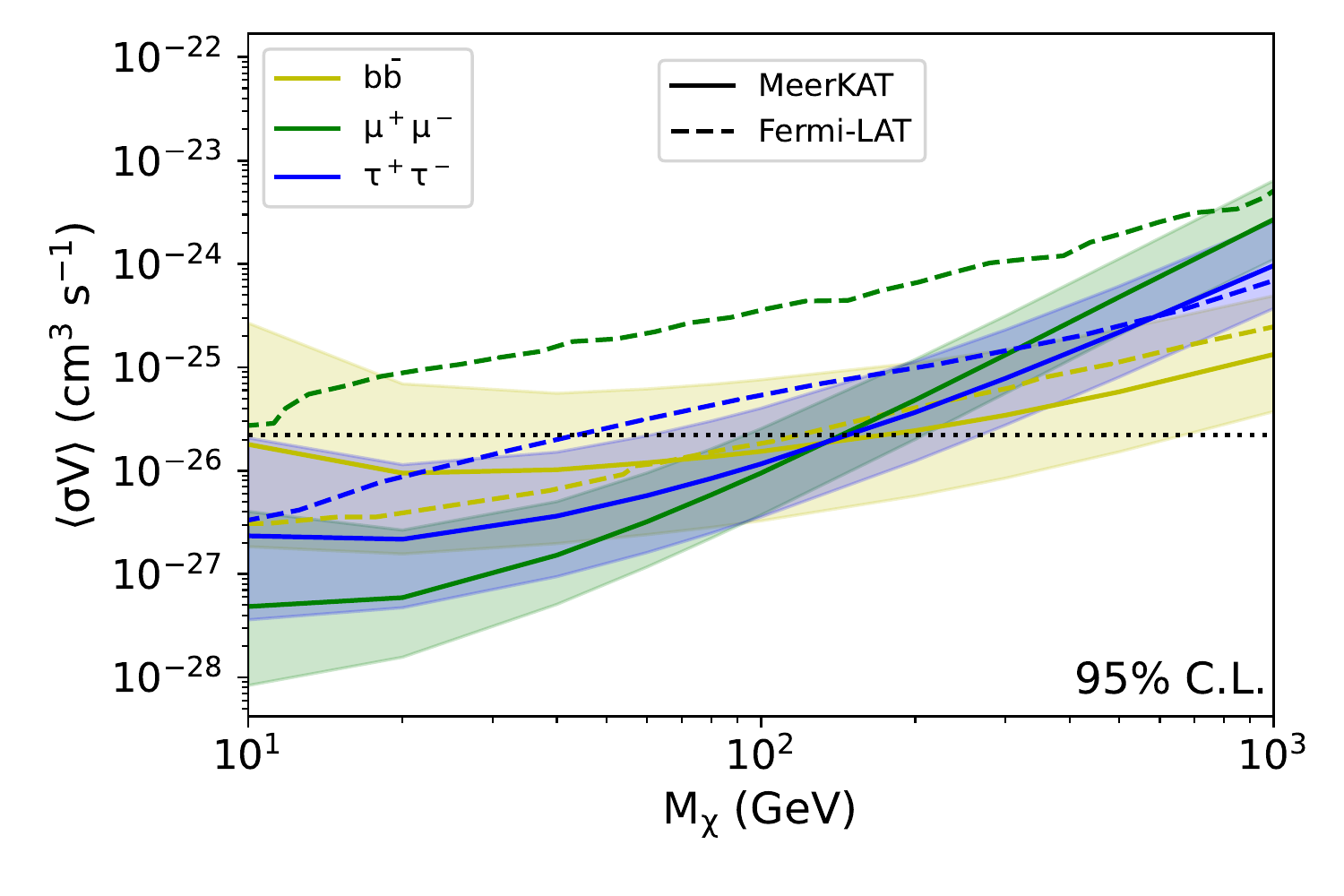}}
    \resizebox{0.48\hsize}{!}{\includegraphics{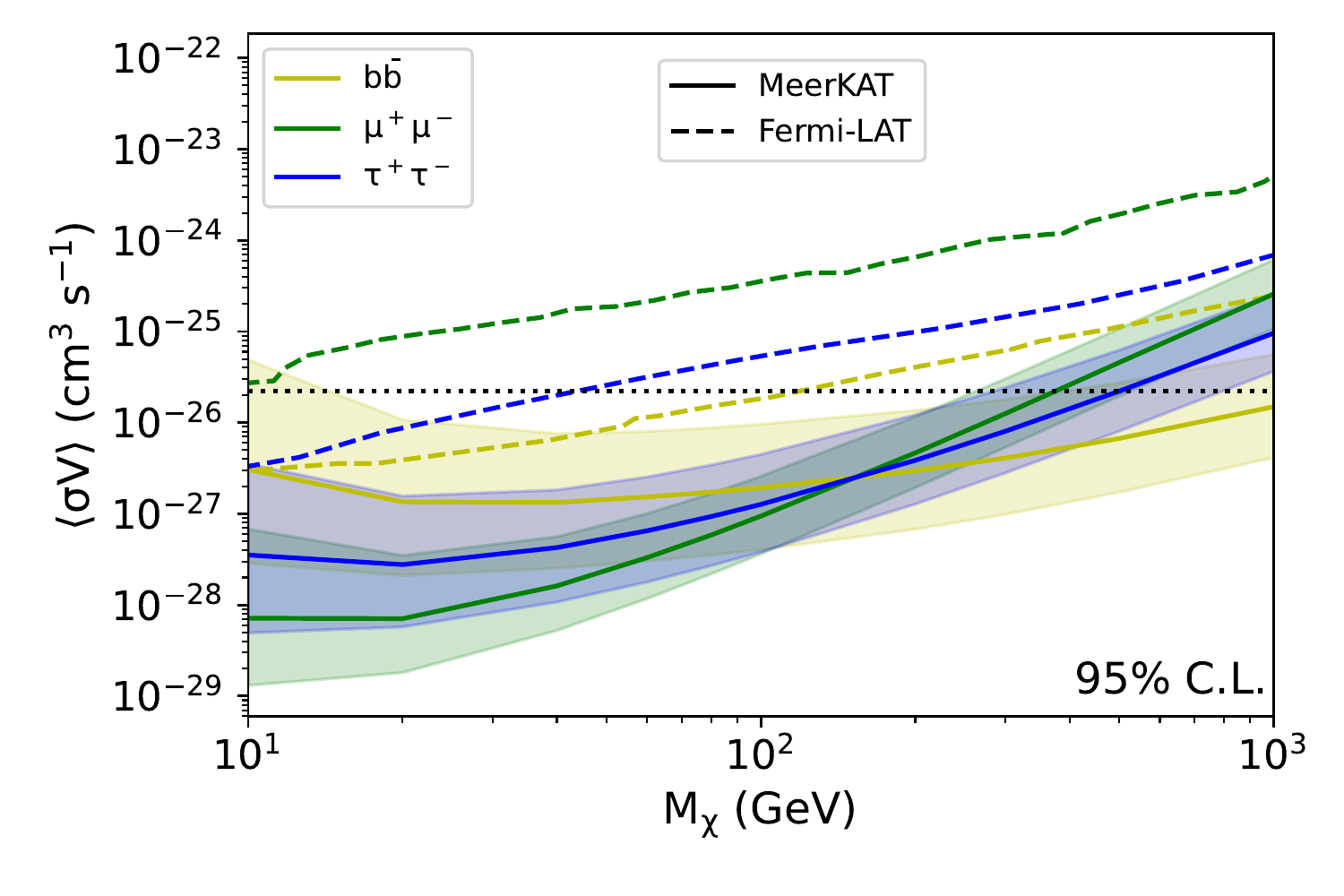}}
    \caption{Projected MeerKAT non-detection constraints for a 20 hour U-band and L-band observation of Reticulum II at the 95 \% confidence interval (solid lines are $B_0 = 1$ $\mu$G, shaded regions show a factor of 2 variance on $B_0$). These projections are contrasted with Fermi-LAT dwarf spheroidal limits from~\cite{Hoof_2020} (dashed lines). The black dotted line is the thermal relic cross-section~\cite{steigman2012}. \textit{Left}: no taper. \textit{Right}: 15$^{\prime\prime}$ Gaussian taper.}
    \label{fig:sb}
\end{figure}

Figure~\ref{fig:sb} shows the potential non-detection limits when considering the surface brightness of Reticulum II. For comparison, we include dwarf galaxy limits from Fermi-LAT~\cite{Hoof_2020} (dashed lines) and shaded regions showing the effect of a factor of 2 uncertainty in the magnetic field strength. It is clear that, even when $B_0 = 0.5$~$\mu$G, the leptonic channels are at least competitive with Fermi-LAT and can greatly exceed it when $B_0 = 1$ $\mu$G (solid lines). These limits are sufficient to rule out WIMP models at the thermal annihilation cross-section~\cite{steigman2012} level for masses below $100$~GeV in all three channels provided $B_0 \approx 1$~$\mu$G. Notably, the magnetic uncertainties could only result in $b$-quark channel being less constraining than Fermi-LAT. 

The use of a taper greatly boosts the constraining power of the instrument, as all three channels exceed Fermi-LAT, even when $B_0 = 0.5$ $\mu$G. In the weakest field scenario we can potentially rule out WIMP annihilation at the thermal relic level for masses up to $\approx 200$~GeV, more than a factor of 2 improvement over the gamma ray experiment. These prospects also improve greatly for the baseline $B_0 = 1$~$\mu$G case, pushing the exclusion to masses below 1~TeV with $b$-quarks, even stronger than the recent powerful limits from a study of the large Magellanic cloud~\cite{Regis_2021}.  

\section{Discussion and conclusions}
\label{sec:disc}

We have shown that the MeerKAT instrument has great potential in the hunt for diffuse synchrotron emission produced by WIMP annihilation. In particular, non-detection limits for the Reticulum II dwarf galaxy exceed the existing Fermi-LAT limits from such galaxies by up to an order of magnitude when making use of a taper on the radio visibilities. These constraints, of course, may not be fully realised, due to the potential to observe weak `baryonic' diffuse emission and will be reduced, in any case, by the need for the extraction of point-source fluxes. On the other hand, the sensitivity estimates used here are likely conservative due to the large channel width being used (for reasons of easier computation). Given that MeerKAT operates with $\sim 200$~kHz channels, this could increase the sensitivity by having fewer channels flagged for radio interference. The main point we wish to highlight is the unexpected potential of the MeerKAT instrument for these searches. We have shown that MeerKAT is far better than ATCA and ASKAP at constraining WIMP dark matter from dwarf galaxies, which is particularly exciting since ATCA and ASKAP have already produced competitive limits~\cite{Regis_2017,Regis_2021}. This suggests that MeerKAT is already a new frontier in indirect DM hunting and is quite emphatically a `Dark Matter Machine', even without the benefit of the full SKA.   

\section*{Acknowledgements}
G.B acknowledges support from the National Research Foundation of South Africa Thuthuka grant (grant no. 117969).
S.M acknowledges support from the South African Radio Astronomy Observatory Salary Research Grant. 

\section*{References}

\bibliographystyle{iopart-num}
\bibliography{refs}

\providecommand{\newblock}{}
\begin{thebibliography}{10}
\expandafter\ifx\csname url\endcsname\relax
  \def\url#1{{\tt #1}}\fi
\expandafter\ifx\csname urlprefix\endcsname\relax\def\urlprefix{URL }\fi
\providecommand{\eprint}[2][]{\url{#2}}

\bibitem{Kenyon_2018}
Kenyon J~S, Smirnov O~M, Grobler T~L and Perkins S~J 2018 {\em Monthly Notices
  of the Royal Astronomical Society\/} {\bf 478} 2399--2415
  \urlprefix\url{https://doi.org/10.1093%2Fmnras%2Fsty1221}

\bibitem{killms1}
{Tasse} C 2014 {\em arXiv e-prints\/} arXiv:1410.8706 (\textit{Preprint}
  \eprint{1410.8706})

\bibitem{killms2}
Tasse C 2014 {\em Astronomy {\&} Astrophysics\/} {\bf 566} A127
  \urlprefix\url{https://doi.org/10.1051%2F0004-6361%2F201423503}

\bibitem{offringa-wsclean-2014}
Offringa A~R, McKinley B, Hurley-Walker {\em et~al.\/} 2014 {\em MNRAS\/} {\bf
  444} 606--619

\bibitem{ddfacet}
Tasse C, Hugo B, Mirmont M, Smirnov O, Atemkeng M, Bester L, Hardcastle M~J,
  Lakhoo R, Perkins S and Shimwell T 2018 {\em Astronomy {\&} Astrophysics\/}
  {\bf 611} A87 \urlprefix\url{https://doi.org/10.1051%2F0004-6361%2F201731474}

\bibitem{pybdsf}
{Mohan} N and {Rafferty} D 2015 {PyBDSF: Python Blob Detection and Source
  Finder} Astrophysics Source Code Library, record ascl:1502.007
  (\textit{Preprint} \eprint{1502.007})

\bibitem{baltz1999}
Baltz E~A and Edsj\"o J 1998 {\em Phys. Rev. D\/} {\bf 59}(2) 023511
  \urlprefix\url{https://link.aps.org/doi/10.1103/PhysRevD.59.023511}

\bibitem{baltz2004}
Baltz E~A and Wai L 2004 {\em Phys. Rev. D\/} {\bf 70}(2) 023512
  \urlprefix\url{https://link.aps.org/doi/10.1103/PhysRevD.70.023512}

\bibitem{Colafrancesco2006}
Colafrancesco S, Profumo S and Ullio P 2006 {\em A\&A\/} {\bf 455} 21

\bibitem{Colafrancesco2007}
Colafrancesco S, Profumo S and Ullio P 2007 {\em Phys. Rev. D\/} {\bf 75}
  023513

\bibitem{siffert2011}
{Siffert} B~B, {Limone} A, {Borriello} E, {Longo} G and {Miele} G 2011 {\em
  Monthly Notices of the Royal Astronomical Society\/} {\bf 410} 2463--2471
  (\textit{Preprint} \eprint{1006.5325})

\bibitem{gsp2015}
Colafrancesco S, Marchegiani P and Beck G 2015 {\em JCAP\/} {\bf 02} 032C

\bibitem{storm2017}
Storm E, Jeltema T~E, Splettstoesser M and Profumo S 2017 {\em The
  Astrophysical Journal\/} {\bf 839} 33
  \urlprefix\url{http://stacks.iop.org/0004-637X/839/i=1/a=33}

\bibitem{natarajan2013}
Natarajan A {\em et~al.\/} 2013 {\em Phys. Rev.\/} {\bf D88} 083535

\bibitem{spekkens2013}
Spekkens K, Mason B~S, Aguirre J~E and Nhan B 2013 {\em Astrophys. J.\/} {\bf
  773} 61

\bibitem{atca_III}
Regis M, Colafrancesco S, Profumo S, de~Blok W, Massardi M and Richter L 2014
  {\em Journal of Cosmology and Astroparticle Physics\/} {\bf 2014} 016
  \urlprefix\url{http://stacks.iop.org/1475-7516/2014/i=10/a=016}

\bibitem{Regis_2017}
Regis M, Richter L and Colafrancesco S 2017 {\em Journal of Cosmology and
  Astroparticle Physics\/} {\bf 2017} 025–025 ISSN 1475-7516
  \urlprefix\url{http://dx.doi.org/10.1088/1475-7516/2017/07/025}

\bibitem{Vollmann_2020}
Vollmann M, Heesen V, Shimwell T~W, Hardcastle M~J, Brüggen M, Sigl G and
  Röttgering H~J~A 2020 {\em Monthly Notices of the Royal Astronomical
  Society\/} {\bf 496} 2663--2672
  \urlprefix\url{https://doi.org/10.1093%2Fmnras%2Fstaa1657}

\bibitem{Basu_2021}
Basu A, Roy N, Choudhuri S, Datta K~K and Sarkar D 2021 {\em Monthly Notices of
  the Royal Astronomical Society\/} {\bf 502} 1605--1611
  \urlprefix\url{https://doi.org/10.1093%2Fmnras%2Fstab120}

\bibitem{Regis_2021}
Regis M, Reynoso-Cordova J, Filipovi{\'{c} } M~D, Brüggen M, Carretti E,
  Collier J, Hopkins A~M, Lenc E, Maio U, Marvil J~R, Norris R~P and Vernstrom
  T 2021 {\em Journal of Cosmology and Astroparticle Physics\/} {\bf 2021} 046
  \urlprefix\url{https://doi.org/10.1088%2F1475-7516%2F2021%2F11%2F046}

\bibitem{fermi}
Albert A, Anderson B, Bechtol K, Drlica-Wagner A, Meyer M, Sánchez-Conde M,
  Strigari L, Wood M, Abbott T~M~C, Abdalla F~B and et~al 2017 {\em The
  Astrophysical Journal\/} {\bf 834} 110 ISSN 1538-4357
  \urlprefix\url{http://dx.doi.org/10.3847/1538-4357/834/2/110}

\bibitem{makhathini2018}
Makhathini S 2018 {\em Advanced radio interferometric simulation and data
  reduction techniques\/} Ph.D. thesis Rhodes University Drosty Rd,
  Grahamstown, 6139, Eastern Cape, South Africa available via
  \url{http://hdl.handle.net/10962/57348}

\bibitem{Pace_2018}
Pace A~B and Strigari L~E 2018 {\em Monthly Notices of the Royal Astronomical
  Society\/} {\bf 482} 3480–3496 ISSN 1365-2966
  \urlprefix\url{http://dx.doi.org/10.1093/mnras/sty2839}

\bibitem{Koposov_2015}
Koposov S~E, Belokurov V, Torrealba G and Evans N~W 2015 {\em The Astrophysical
  Journal\/} {\bf 805} 130 ISSN 1538-4357
  \urlprefix\url{http://dx.doi.org/10.1088/0004-637X/805/2/130}

\bibitem{steigman2012}
Steigman G, Dasgupta B and Beacom J~F 2012 {\em Phys. Rev.\/} {\bf D86} 023506
  (\textit{Preprint} \eprint{1204.3622})

\bibitem{atca_II}
Regis M, Richter L, Colafrancesco S, Profumo S, de~Blok W~J~G and Massardi M
  2015 {\em Monthly Notices of the Royal Astronomical Society\/} {\bf 448}
  3747--3765 \urlprefix\url{https://doi.org/10.1093%2Fmnras%2Fstv127}

\bibitem{rybicki1986}
{Rybicki} G~B and {Lightman} A~P 1986 {\em {Radiative Processes in
  Astrophysics}\/} (Wiley)

\bibitem{longair1994}
Longair M~S 1994 {\em High Energy Astrophysics\/} (Cambridge University Press)

\bibitem{einasto1968}
Einasto J 1968 {\em Publications of the Tartuskoj Astrofizica Observatory\/}
  {\bf 36} 414

\bibitem{walker2009}
Walker M~G, Mateo M, Olszewski E~W, narrubia J~P, Evans N~W and Gilmore G 2009
  {\em ApJ\/} {\bf 704} 1274
  \urlprefix\url{http://stacks.iop.org/0004-637X/704/i=2/a=1274}

\bibitem{adams2014}
Adams J~J {\em et~al.\/} 2014 {\em ApJ\/} {\bf 789} 63
  \urlprefix\url{http://stacks.iop.org/0004-637X/789/i=1/a=63}

\bibitem{casa}
{McMullin} J~P, {Waters} B, {Schiebel} D, {Young} W and {Golap} K 2007 {\em
  Astronomical Data Analysis Software and Systems XVI\/} ({\em Astronomical
  Society of the Pacific Conference Series\/} vol 376) ed {Shaw} R~A, {Hill} F
  and {Bell} D~J p 127

\bibitem{meqtrees}
{Noordam, J E} and {Smirnov, O M} 2010 {\em A\&A\/} {\bf 524} A61
  \urlprefix\url{https://doi.org/10.1051/0004-6361/201015013}

\bibitem{Hoof_2020}
Hoof S, Geringer-Sameth A and Trotta R 2020 {\em Journal of Cosmology and
  Astroparticle Physics\/} {\bf 2020} 012--012
  \urlprefix\url{https://doi.org/10.1088%2F1475-7516%2F2020%2F02%2F012}

\end{thebibliography}

\end{document}